\documentclass[aps,twocolumn,showpacs,preprintnumbers,superscriptaddress,amsmath,amssymb]{revtex4}

\usepackage{graphicx}
\usepackage{dcolumn}
\usepackage{bm}

\begin{document}

\title{Tunneling into Multiwalled Carbon Nanotubes: Coulomb Blockade and Fano Resonance\\}

\author{W.  Yi}
\altaffiliation[Present address: Gordon McKay Laboratory of
Applied Science, Harvard University, MA 02138, USA]{}

\author{L. Lu}
\affiliation{Institute of Physics, Chinese Academy of Sciences,
Beijing 100080, P. R. China}

\author{H. Hu}
\affiliation{Department of Physics, Tsinghua University, Beijing
100084, P. R. China}

\author{Z. W. Pan}
\affiliation{Institute of Physics, Chinese Academy of Sciences,
Beijing 100080, P. R. China}

\author{S. S. Xie}
\affiliation{Institute of Physics, Chinese Academy of Sciences,
Beijing 100080, P. R. China}


\begin{abstract}
Tunneling spectroscopy measurements of single tunnel junctions
formed between multiwalled carbon nanotubes (MWNTs) and a normal
metal are reported. Intrinsic Coulomb interactions in the MWNTs
give rise to a strong zero-bias suppression of a tunneling density
of states (TDOS) that can be fitted numerically to the
environmental quantum-fluctuation (EQF) theory.  An asymmetric
conductance anomaly near zero bias is found at low temperatures
and interpreted as Fano resonance in the strong tunneling regime.
\end{abstract}

\pacs{73.63.Fg, 73.23.Hk, 85.35.Kt}

\maketitle

Coulomb blockade (CB) has been studied intensely in a
multi-junction configuration, in which electron tunnel rates from
the environment to a capacitively isolated ``island" are blocked
by the e-e interaction if the thermal fluctuation is below the
charging energy $E_c = e^{2}/2C$ and the quantum fluctuation is
suppressed with sufficiently large tunnel resistance $R_t \gg R_Q
= h^{2}/2e$. In the case of a single-junction circuit, the
understanding of CB is less straightforward. The Coulomb gap,
supposedly less significant for the case of low-impedance
environments, should be established only if the environmental
impendence exceeds $R_Q$.

In single-walled carbon nanotubes (SWNTs), the reduced geometry
gives rise to strong e-e interaction. Indeed, CB oscillations and
evidence of Luttinger liquid (LL) have been observed
\cite{Bockrath}. In contrast to SWNTs, in which only two
conductance channels are available for current transport, MWNTs
with diameter in the range of $d$=20-40 nm have several tens of
conductance channels. The energy separation of the quantized
subbands, given by $\Delta E = \hbar \upsilon_f/d$, is about 13-26
meV taking the Fermi velocity $\upsilon_f = 8 \times 10^{5}$ m/s.
This value is about an order of magnitude smaller than that of
SWNTs. Experiments indicate that MWNTs are considerably
hole-doped, thus a large number of subbands, on the order of ten,
are occupied. Observations of weak localization \cite{Langer},
electron phase interference effect \cite{Bachtold2} and universal
conductance fluctuations \cite{Langer} support the view that low
frequency conductance in MWNTs is contributed mostly by the
outmost graphene shell and is characterized by 2D diffusive
transport. In addition to the phase interference effects, a strong
e-e interaction has also been observed in MWNTs. Pronounced
zero-bias suppression of the TDOS has been observed several times
in the tunneling measurements \cite{Haruyama, Tarkiainen,
Bachtold}. Moreover, the TDOS shows a power-law, i.e., $\nu(E)
\sim E^{\alpha}$, which resembles the case of a LL. It is
noteworthy that in the EQF theory, for a single tunnel junction
coupled to high-impedance transmission lines, such a scaling
behavior is also predicted at the limit of many parallel
transmission modes \cite{Ingold}. The physical origin of these
power laws is the linear dispersion of bosonic excitations that
are characteristic both for LL, which is a strictly 1D ballistic
conductor, and a single tunnel junction connected to a 3D
disordered conductor. In the latter case, the quasiparticle
tunneling is suppressed at $V \ll e/2C$, therefore the charge is
transported with 1D plasmon modes. The fact that MWNTs have many
conductance modes, together with the observation of a crossover
from power-law to Ohmic behavior at higher voltages
\cite{Tarkiainen}, suggests that the EQF theory is more
appropriate to describe the observed TDOS renormalization.
However, most of these measurements were done in multi-junction
configurations. A single tunnel junction measurement is needed to
further clarify this issue.

In this Letter, millimeter-long CVD-grown MWNTs \cite{Pan} are
measured by a cross-junction method. The MWNTs samples are
composed of loosely entangled nanotubes of diameters between 20 to
40 nm that are roughly parallel to each other and up to 2 mm long.
The single tunnel junctions are formed by crossing a very thin ($<
1 \mu$m) MWNT bundle with a narrow strip of metal wire fabricated
on an insulating substrate. We explored different electrode
materials including Au, Cu, Sn, and Al. In the case of Sn and Al,
a small magnetic field is applied to suppress the superconducting
state below $T_c$. No obvious change of device characteristics
attributable to the choice of metals is observed. With such a
cross-junction configuration the measured conductance is
contributed exclusively by the tunnel junction, since the current
is passed along one arm of the MWNT/metal and the voltage is
measured along the other, non-current carrying arm. Thus the
device can be understood as a small number of single junctions in
parallel. Despite the simplicity of their fabrication, we find
that the devices are very stable and sustain several cooling
cycles without apparent change of characteristics. More than 20
samples are measured, all yielding strong zero-bias suppression of
the TDOS.

\begin{figure}
\includegraphics[width = 8.2cm, height = 7cm]{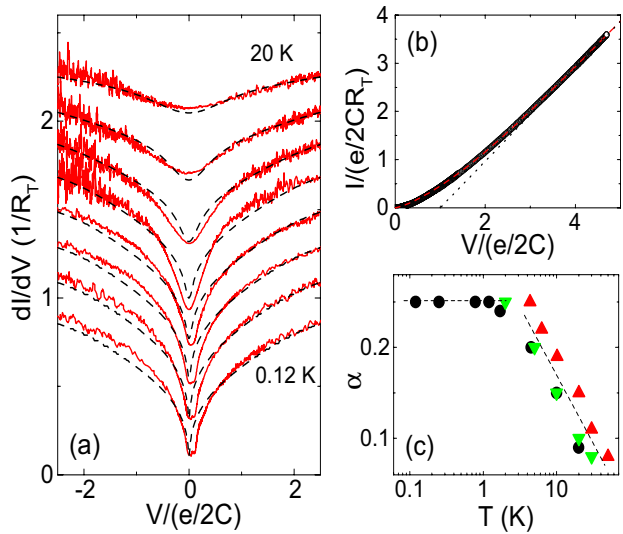}
\caption{\label{fig1} (a) $dI/dV$ as a function of $V$ in
dimensionless units measured at $T$ = 0.12, 0.25, 0.78, 1.2, 1.7,
4.5, 10, 20 K (sample 990530s6, curves are offset for clarity).
Dashed lines are the fits of EQF theory. (b) DC current
simultaneously measured at 0.12 K. Dashed line is the numerical
fit. Dotted line shows a Coulomb offset of $e/2C$. (c) The
temperature dependence of the exponent $\alpha$ for three samples
($\circ$: 990530s6; $\bigtriangleup$: 990316; $\bigtriangledown$:
990320).}
\end{figure}

The current $I(V)$ and $G \equiv dI/dV$ are calculated by a
golden-rule approach incorporating the environmental influence by
$P(E)$, the probability for a tunneling electron to lose energy to
the environment. $P(E)$ can be calculated from its Fourier
transform, the phase correlation function $J(t)$. In the
transmission-line model, $J(t)$ is determined by the total
environmental impedance $Z_t (\omega) = (i\omega C +
Z^{-1}(\omega))^{-1}$, where $Z(\omega)$ is the external
environmental impedance. We apply the method in Ref.
\cite{Ingold2} to evaluate $P(E)$ from an integral equation
without the need of going to the time domain. The I-V
characteristics are then calculated for finite temperatures and
arbitrary junction impedance. The parameters that appear in the
numerical calculation are the damping strength $\alpha =
Z(0)/R_Q$, the inverse relaxation time $\omega_{RC}$, and the
quality factor $Q = \omega_{RC} / \omega_S$ with the resonance
frequency of the undamped circuit $\omega_S = (LC)^{-1/2}$. It is
noted that the quality factor $Q$ plays only a minor role as an
additional adjustable parameter. Therefore it is not important to
the fit, and is always set to unit.

We then try to fit our experimental data with the EQF theory.
Since $E_c$ and $R_t$ are determined by the high-voltage data, the
isolation resistance $R_{iso} = Z(0)$ is the only adjustable
parameter in our fitting. In contrast to the case of Ref.
\cite{Zheng}, where $R_{iso}$ is formed by ideal Ohmic and
temperature-independent resistors, $R_{iso}$ in our case are
provided by the resistive impedance of the MWNTs themselves, which
should be temperature-dependent. Indeed, from the fitting, we see
an evolution of $\alpha$ from 0.1 at 20 K to 0.25 at 1-4 K and
then $\alpha$ saturates (Fig. 1c). Note that dimensionless units
are used with $G$ normalized by 1/$R_t$ and voltage normalized by
$e/2C$. Therefore the number of the single tunnel junctions has no
effect on the fitting, and our measurements of different samples
can be directly compared. We find that for different samples with
scattered characteristics (see Table I), at low temperatures the
exponent $\alpha$ reaches a universal value of 0.25-0.35, which
agrees with previous multi-junction measurements \cite{Tarkiainen,
Bachtold}. It yields $R_{iso}$ = 3.3-4.6 k$\Omega$, which is
roughly a constant for different samples. This coincidence is not
accidental but reflects the intrinsic electrodynamic modes of the
MWNTs, which can be modelled as an ideal resistive LC transmission
line: $R = (L'/C')^{1/2}$ with the kinetic inductance estimated as
$L' = R_Q/2N\upsilon_f \approx 1$ nH/$\mu$m for $N \approx$ 10 -
20 modes and the capacitance $C' \approx$ 20-30 aF/$\mu$m.
Therefore, the ``environment" with respect to the single-junctions
in our devices is provided by MWNTs themselves, not by the
external circuits.

\begin{table}
\caption{\label{tab:table1}A partial list of characteristics of
the samples.}
\begin{ruledtabular}
\begin{tabular}{cccccccc}
Sample &$R_t$ (k$\Omega$)&$E_c$ (eV) &$C$ (aF) &$R_{iso}$ (k$\Omega$)&$\alpha$\\
\hline
990316& 18.4 & 0.019 & 4.2 &3.3\footnotemark[1] & 0.25\footnotemark[1] \\
990320& 48.3 & 0.014 & 5.7 &3.3\footnotemark[1] & 0.25\footnotemark[1] \\
990530s6& 8.6 & 0.01 & 8.0 &3.3\footnotemark[1] & 0.25\footnotemark[1] \\
990530s1& 10.0 & 0.02 & 4.0 &4.6\footnotemark[1] & 0.35\footnotemark[1] \\
990202& 9.1 & 0.004 & 20.0 &3.5\footnotemark[1] & 0.27\footnotemark[1] \\
\end{tabular}
\end{ruledtabular}
\footnotetext[1]{value acquired at $T$ = 1.2--4.3 K.}
\end{table}

As mentioned previously, for an isolated single tunnel junction
with many parallel transport modes, the EQF theory predicts a
power-law asymptotics: $G(V,T)/G(0,T) \equiv f(V/T) = \lvert
\Gamma[\frac{1} {2} (\alpha +2) + \frac{ieV}{2\pi
k_BT}]/\Gamma[\frac{1} {2} (\alpha +2)]\Gamma[1 + \frac{ieV}{2\pi
k_BT}]\rvert^{2}$, where $G(0,T)$ is the zero-bias conductance,
and $\Gamma (x)$ is the Gamma function \cite{Zheng}. For $eV/2\pi
k_B T \gg 1$, a voltage power law $G(V,T) \sim V^\alpha$ is
expected. Note that in the above scaling function, exponent
$\alpha$ is the only adjustable parameter. In Fig. 2, we can
observe such scaling behavior. Here $f(V/T)$, represented by solid
line, is calculated taking $\alpha = 0.27$. The inset shows the
original data in which the dashed line is calculated numerically
using the same exponent $\alpha = 0.27$.

CB in MWNTs is recently described with a microscopic theory in
Ref. \cite{Egger}. In the 2D diffusive regime, the exponent for
tunneling into the bulk of a MWNT is: $\alpha
=(R/h\nu_{0}D)ln(1+\nu_{0}U_{0})$. Here $R$ is the tube radius,
$U_0$ is the intra-tube Coulomb interaction, $D = \upsilon^{2}_f
\tau/2$ is the charge diffusivity and the ``bare" DOS $\nu_0 =
N/4h\upsilon_f$ with $N \approx 20$. If we use $\upsilon_f\tau
\approx 60$ nm from the magnetoresistance data \cite{KangMR}, then
under the condition of $U_0/h\upsilon_f \sim 1$ the above formula
yields $\alpha \approx 0.25$, which agrees with our experiment.

\begin{figure}
\includegraphics[width = 6cm, height = 6.3cm]{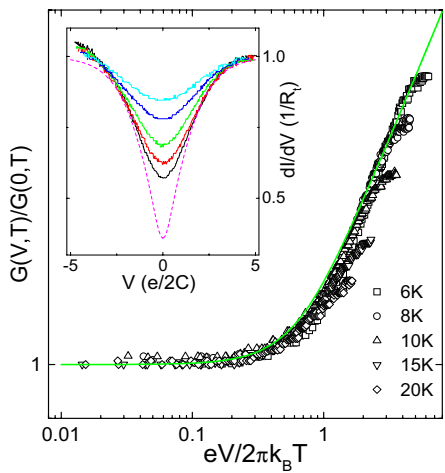}
\caption{\label{fig2} Scaled conductance $G(V,T)/G(0,T)$ of
another sample (990202). The inset shows the original $dI/dV vs.
V$ with the dashed line calculated with exponent $\alpha = 0.27$.}
\end{figure}

Besides the voltage power law phenomenon, which is characteristic
of strong Coulomb interactions, another energy scale - the
discrete energy levels due to electron confinement - emerges as
temperature decreases, and it modifies the tunneling spectra.
Unlike the case of a two-junction configuration, where the
electrons are confined by the two contacts if the nanotube is
clean enough, in a single-junction configuration the electrons can
be confined by disorders when the MWNTs are ``dirty", such that
the impedance of the local environment of the junction is larger
than $R_Q$. A stacking mismatch between adjacent walls and other
structural imperfections are possible sources of disorders in
MWNTs, resulting in discrete energy levels.

By cooling down the devices to below 1 K, we observe that $G$
develops a narrow resonance-like anomaly at very low bias (Fig.
3). The asymmetric anomaly builds up consistently as temperature
decreases and even shows a dip structure. The line shape resembles
that of a Fano resonance.

Unlike the Coulomb structure, which is caused by static e-e
interaction, a Fano resonance arises from an e-e exchange
interaction between two interfering scattering channels: a
discrete energy level and a continuum band. It has recently been
rediscovered in mesoscopic systems such as semiconductor quantum
dots, SWNTs, etc. \cite{Gores, Wiel, Nygard, Liang}. Taking a
quantum dot as an example, it can be considered as a gate-confined
droplet of electrons with localized states. The coupling of the
dot to the leads can be tuned to control the system to enter
different transport regimes: If the dot is weakly coupled with the
environment, a well-established CB develops. The charge transport
is suppressed except for narrow resonances at charge degeneracy
points. When the tunnel barriers become more transparent, the dot
enters the Kondo regime -- i.e., below a characteristic Kondo
temperature $T_K$, spin-flip co-tunneling events introduce a
narrow symmetric TDOS peak at $E_F$ that can be interpreted as a
discrete level. If the coupling is strong enough, the interference
between this discrete level and the conduction continuum gives
rise to an asymmetric Fano resonance. In Ref. \cite{Gores}, such a
crossover from a well-established CB through Kondo regime to Fano
regime has been clearly observed.

\begin{figure}
\includegraphics[width = 8.5cm, height = 4.5cm]{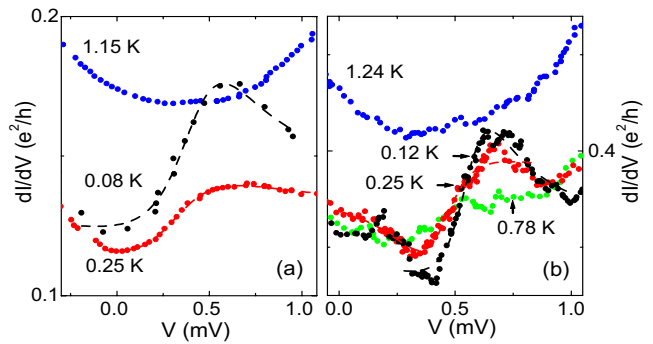}
\caption{\label{fig3} Asymmetric resonance features of $dI/dV$
seen in two samples: 990530s1 (a) and 990530s6 (b). Dashed lines
are the fits of Fano's formula.}
\end{figure}

\begin{table}
\caption{\label{tab:table2}Characteristics of the two samples in
Figure 3.}
\begin{ruledtabular}
\begin{tabular}{cccccccc}
Sample &$T$ (K) &$q$ &$\varepsilon_0$ (meV) &$\gamma$ (meV) &$T_K$ (K)\\
\hline
990530s6& 0.25 & 0.53 & 0.43 & 0.32 & 1.86\\
990530s6& 0.12 & 1.75 & 0.57 & 0.29 & 1.68\\
990530s1& 0.25 & 0.92 & 0.3 & 0.65 & 3.77\\
990530s1& 0.08 & 1.97 & 0.43 & 0.56 & 3.25\\
\end{tabular}
\end{ruledtabular}
\end{table}

We find that the asymmetric resonance curve of $G$ can be fitted
by the Fano's formula: $G \sim (\epsilon + q)^{2}/(\epsilon^{2} +
1)$. Here $q$ is the so-called asymmetry parameter, $\epsilon =
(eV - \varepsilon_0)/(\gamma/2)$ is the dimensionless detuning
from resonance, and $\gamma$ is the FWHM of the resonance. As
expected, the asymmetric parameter $q$ as a measure of the degree
of coupling between the discrete state and the continuum increases
when the temperature drops.

The Kondo temperature, estimated from the relation $\gamma =
2k_BT_K$, is consistent for each device at different temperatures,
and agrees with the observed FWHM of the resonance. If Kondo
physics truly exists, then $G$ at voltages near the resonance
should show non-monotonic temperature dependence around $T_K$
\cite{Nygard}. Indeed, the $G-V$ curves in Fig. 3 exhibit such
behavior: $G$ at the peak position first drops with $T$ and then
rises up below $\sim 1$ K (plotted in Fig. 4b). Moreover, we find
that a perpendicular magnetic field gives rise to effects that are
two-fold (Fig. 4a and 4c): First, the background conductance
increases monotonically with the applied $B$. Second, the dip seen
at zero field gradually disappears and the resonance turns into a
nearly symmetric peak at high field, similar to the pattern
observed in semiconductor quantum dots \cite{Gores}.

The magnetic field effect can be explained as follows: First,
adding a flux should change the amplitude and/or the phase for the
resonant channels and therefore break down the coherent
backscattering and increase the forward transmission through the
channel. Second, the magnetic field can destroy the interference
between the resonant and nonresonant paths, transforming a
resonant dip into a peak.

\begin{figure}
\includegraphics[width = 8.5cm, height = 7cm]{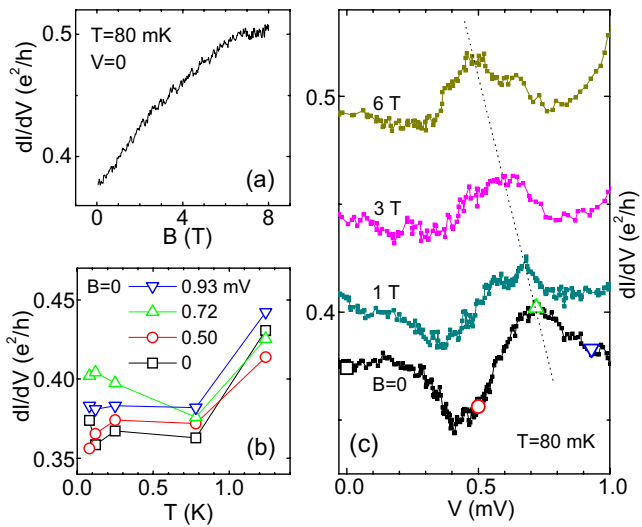}
\caption{\label{fig4} (a) The dependence of zero-bias $dI/dV$ on
perpendicular magnetic field. (b) The temperature dependence of
$dI/dV$ measured at different voltages. (c) Effect of
perpendicular magnetic field. From bottom to top: $B = 0, 1, 3, 6$
Tesla (sample 990530s6).}
\end{figure}

Since Kondo physics is historically interpreted as the interplay
between the d-orbitals of magnetic impurities and the conduction
continuum, we have to eliminate the possibility that the observed
Fano resonance comes from residual traces of the Fe/Si catalyst in
the MWNT samples. It has been shown that magnetic impurities in
MWNTs cause an enhancement of thermoelectric power
\cite{Grigorian}, which is absent in our careful TEP measurements
\cite{KangTEP}. Furthermore, no Fe signature can be detected in
the body of the MWNT bundles within the instrumental resolution in
the energy dispersion X-ray and TEM studies. Therefore the
observed Fano resonance must be an inherent property of MWNTs,
which can be virtually treated as quantum dots strongly coupled to
the leads. We noticed that recently Buitelaar et al. have also
seen clear traces of CB and Kondo resonance in MWNT SET devices
\cite{Buitelaar}. Their observed Kondo temperature $T_K = 1.2$ K
is in good agreement with our results.

In summary, strong zero-bias suppression of the TDOS is observed
in MWNTs' single tunnel junctions that can be explained well by
the EQF theory. The observed exponent $\alpha \approx$ 0.25--0.35
is found to be consistent for all the samples. Similar to the case
of an open quantum dot, in low-impedance junctions we find that a
Fano-resonance-like asymmetric conductance anomaly builts up below
mV energy scales. It seems that MWNTs provide us a good laboratory
to study the interplay between strong e-e interaction and disorder
scattering.

The authors acknowledge fruitful discussions with R. Egger, M.
Bockrath, W. Zheng, T. Xiang and G.M. Zhang. This work is
supported by the National Key Project for Basic Research, the
Knowledge Innovation Program of CAS, and NSFC.


\begin{thebibliography}{20}

\bibitem{Bockrath} M. Bockrath, {\it et al.}, {\it Nature} {\bf 397}, 598 (1999).

\bibitem{Bachtold2} A. Bachtold, {\it et al.}, {\it Nature} {\bf 397}, 673 (1999).

\bibitem{Langer} L. Langer, {\it et al.}, {\it Phys. Rev. Lett.} {\bf 76}, 479 (1996).

\bibitem{Haruyama} J. Haruyama, I. Takesue, and Y. Sato, {\it Appl. Phys. Lett.} {\bf 77}, 2891 (2000).

\bibitem{Bachtold} A. Bachtold, {\it et al.}, {\it Phys. Rev. Lett.} {\bf 87}, 166801 (2001).

\bibitem{Tarkiainen} R. Tarkiainen, {\it et al.}, {\it Phys. Rev. B} {\bf 64}, 195412 (2001).

\bibitem{Ingold} G. -L. Ingold and Y. V. Nazarov, in {\it Single Charge Tunneling}, edited by
H. Grabert and M. H. Devoret, NATO ASI Series (Plenum, New York, 1991).

\bibitem{Pan} Z. W. Pan, {\it et al.}, {\it Nature} {\bf 394}, 631 (1998).

\bibitem{Ingold2} G.-L.Ingold and H. Grabert, {it Europhys. Lett.} {\bf 14}, 371 (1991).

\bibitem{Zheng} W. Zheng, J. R. Friedman, D. V. Averin, S. Han and J. E. Lukens, {\it Solid State Comm.}
{\bf 108}, 839 (1998).

\bibitem{Egger} R. Egger, A. O. Gogolin, {\it Phys. Rev. Lett.} {\bf 87}, 66401 (2001).

\bibitem{KangMR} N. Kang, {\it et al.}, {\it Phys. Rev. B} {\bf 66}, 241403(R) (2002).

\bibitem{Gores} J. Gores, {\it et al.}, {\it Phys. Rev. B} {\bf 62}, 2188 (2000).

\bibitem{Wiel} W. G. van der Wiel, {\it et al.}, {\it Science} {\bf 289}, 2105 (2000).

\bibitem{Nygard} J. Nygard, D. H. Cobden, and P. E. Lindelof, {\it Nature} {\bf 408}, 342 (2000).

\bibitem{Liang} W. Liang, M. Bochrath, H. Park, {\it Phys. Rev. Lett.} {\bf 88}, 126801 (2002).

\bibitem{Grigorian} L. Grigorian, {\it et al.}, {\it Phys. Rev. B} {\bf 60}, R11309 (1999).

\bibitem{KangTEP} N. Kang, {\it et al.}, {\it Phys. Rev. B} {\bf 67}, 033404 (2003).

\bibitem{Buitelaar} M. R. Buitelaar, A. Bachtold, T. Nussbaumer, M. Iqbal, and C. Schonenberger,
{\it Phys. Rev. Lett.} {\bf 88}, 156801 (2002).

\end{thebibliography}
\end{document}